
\magnification=1200
\rightline{RU--95--79}
\def\sla{\raise.15ex\hbox{$/$}\kern-.57em}
\vskip 3cm
\centerline {\bf {A Lecture on Chiral Fermions.}}
\vskip 2cm
\centerline{ Herbert Neuberger}
\vskip 1cm
\centerline{\it {Department of
Physics and Astronomy, Rutgers University,
Piscataway NJ08855, USA}}
\vfill
\noindent{\bf Abstract:} { This is an informal and
approximate transcription of
a talk presented at the DESY workshop,
September 27--29, 1995. The basic
message is that real and long overdue progress is
taking place on the problem of regulating
non--perturbatively chiral gauge theories.
Several approaches are reviewed
with emphasis on the overlap and some of the
questions raised about it. No claim for
completeness or objectivity
is made.}
\vfill\eject
My talk addresses the following objective:
{\bf Construct a non--perturbative
definition of asymptotically free chiral gauge theories in four dimensions.}
The reason for trying to do this is
that fermions in Nature are Weyl, not
Dirac.

I assume everybody here has heard that many
attempts to attain the objective
have failed in the past. The last few annual lattice
conferences had plenary talks [1] which offered no
realistic expectations for concrete progress. A notable
exception to this defeatist mood
was the Rome group [2], which for a while
was quite hopeful. But, the
norm of the activity in the field has been
to explain how something went
wrong for some very technical reasons.
My purpose today is to focus on what
I perceive as some real, long overdue,
progress. I believe that in the not too
far future the basic view will reverse itself
in that the objective will
be viewed as attainable, albeit with a large
(however not prohibitive)
expense of computer time. Today there still
are theoretical physicists (Banks [3], Casher [3],
Friedan [4], Nielsen [5]) tentatively
imputing lattice difficulties
to some fundamental reason (which would always
prevent us from achieving the objective in the
context of field theory).
Some view this as evidence
that the fundamental laws of physics
are to be found in string theory [4]. I'll try to
convince you that nothing as drastic is
needed in order to realize
our relatively modest objective.
Strings may be necessary to satisfy
some needs, but not that for chiral fermions.

First there is a question of principle: is it possible at all
to attain the objective, or
else an asymptotically free
chiral gauge theory, unlike a vector one,
will always have some small
indeterminacy in it, reflecting the larger
theory into which it is embedded,
no matter how far the new scale is set.
In other words, the new scale cannot be
removed completely, a situation similar
(but differing in mechanism) to the
case of a ``trivial'' theory (e.g. $\lambda\phi^4$).
Second comes a question of practice:
How can we compute non--perturbatively
in an asymptotically free chiral gauge theory ?
My talk addresses both questions. In a nutshell
my message will be that progress has been
attained by recognizing that the
number of fermions one needs
to keep in the regularized theory
is still infinite. Several
different ways of doing this have
been explored already and many more are
possible. From this point of view, the
previous attempts that failed did so because they
had a finite number of fermions at the IR and UV regulated level.

The recent progress was initiated by a paper
by Kaplan [6] who exploited earlier work in a different context by
Callan and Harvey [7]. An independent article, by
Frolov and Slavnov [8], appeared at roughly the same time and
suggested a totally different approach which subsequently
turned out to be equally important.
Narayanan and I [9] synthesized the new ideas in these two papers
into one principle of having an infinite number of
flavors. This principle is more evident in the Frolov--Slavnov
approach than in the Kaplan one, but is the basic mechanism
for both schemes, and can be implemented in new ways, not all
yet explored. More recently,
and independently, 't Hooft [10] stressed the same
point, but elected to
have an infinite number of fermions by
keeping them in the continuum,
i.e. never putting them
on the lattice. This has occurred to some people before
(Flume, Wyler [11], and others [12]); what is new
is the view that the need to keep an indefinite
number of fermionic
degrees of freedom at the regulated level
is inevitable because it is
a {\it dynamical} variable dependent on
background gauge field topology.

With the benefit of hindsight we can identify other indications
that chiral fermion systems should not be truncated to
a finite number of Grassmann degrees of freedom per unit volume:

The first indication
I came across was when dealing with
two dimensional bosonization [13].
If you can bosonize a chiral theory and regulate
the bosonic version (maybe with lots
of fine tuning, but still possible)
how come you are having trouble at the fermionic level ?
I knew then that the answer
was that the regulated quantum bosonic space still was
infinite dimensional
and a traditionally
regulated fermionic theory lived in a finite
dimensional Hilbert space (for fixed values of
the original bosonic degrees of freedom).
So the bosonic regularization
could not be ``fermionized''.
I now suspect that
UV regulated bosonization
is possible if one keeps an
infinite number of fermionic species.

Nielsen's [14] physical analysis of
the source of the anomaly in the presence
of constant external fields showed
that an ``infinite hotel'' was
active in the chiral case,
but that it could be truncated and
``piped'' into another if the
theory was vectorial.
He also argued that this
``plumbing'' would not work in the
chiral case even if anomaly
free and thus the latter is doomed.
The new trick is to
house each multiplet in its own infinite hotel.

The most direct evidence
for the need to keep the number of fermions
infinite comes from the
Atiyah -- Singer theorem [15] and the 't Hooft vertices [16]
associated with it. The
Weyl operator, in the Euclidean context, when
defined on a compact base
manifold to eliminate infrared problems,
has a nontrivial index.
This means that the operator has more
(or less) zero modes than its hermitian conjugate. Traditional
non--perturbative  regularizations always replace the Weyl operator
by some finite, square matrix of fixed, (albeit huge) size. But
the square matrix
cannot have an index and,
as a result, 't Hooft vertices typically
get zero expectation value.
Using finite square matrices for representing the
chiral Weyl operator is analogous to trying
to study the harmonic oscillator replacing $a$ and $a^\dagger$ by
finite square
matrices. All goes well, except
the commutation relation is badly mangled
since its trace must vanish now.
If you ask questions about some very
low states
you may see no problem, but the uncertainty relation for
$a\pm a^\dagger$ does not hold
and you must be careful with some
conclusions your cutoff theory may lead you to.

Starting from the point that the number of fermionic
degrees of freedom interacting with a lattice field cannot be
fixed 't Hooft decided to keep the fermions in the continuum and
consider their interaction with a continuum connection derived
from a lattice gauge field configuration. The integration over
the fermions is a super--renormalizable problem, so it can be
regulated by the addition of a few Pauli Villars regulator fields
and the counter--terms can be exactly computed.
Gauge invariance is broken at the regulated level but is restored
when the Pauli--Villars masses are taken to infinity if the
theory is anomaly free. The main problem is to provide
a non--perturbative definition of the fermion induced determinant
that is good enough to work for all continuum connections
one could conceivably construct from the lattice link variables.
If the continuum connection carries nontrivial topology the
integral over the fermions yields
zero, and this is it. There remains something to be done
only for zero topology.

Frolov and Slavnov started with SO(10) and a single 16 in 4D where
UV infinities occur only in the parity conserving parts
of the effective chiral action. But they come in with
one half the pre--factor of the vectorial case.
Taking a square root is a
nonlocal operation, so one would like to put in wrong
statistics fields to cancel
the unwanted half of the coefficient.
However, as long as you add a finite
number of Dirac fields, obviously, only
integers can be canceled out.
But, sometimes 1-1+1-1.... can be 1/2
if this keeps on going ad infinitum.
They went ahead and invented a set of
masses for an infinite number of
fermions of correct and opposite
statistics that did the job.

Kaplan started from a paper by
Callan and Harvey who set up a physical
model where the strange relation
between the anomaly functionals in
different dimensions (Stora, Zumino [17])
could be understood. They dealt with both the d to
d-2 and d to d-1 relation, but Kaplan singled out the
case with odd d and the relation to d-1 (even).
Callan and Harvey showed that in a five dimensional
theory with a mass defect a
massless and chiral mode appeared bound to
the defect. The effective
action describing the four dimensional world
at the defect had to have no
local charge conservation due to the anomaly,
but in the five dimensional
setup charge was conserved. The resolution to
the puzzle was that the five dimensional
current carried off to the infinities
in the directions perpendicular to the
defect a net amount of charge making
the myopic four dimensional world conclude
that charge is not conserved.
The well--known relation between the five
dimensional Chern Simons terms
(giving the
asymptotic form of the Goldstone--Wilczek [18] currents
far away from the defect, but induced by it)
and the four dimensional anomaly
showed that the lack of non--conservation
was of the correct form.
Kaplan realized that the entire
Callan Harvey setup could be put on
the lattice without any trouble
(this depended on an extension of the Wilson doubler removal
mechanism to the odd dimensional case and on the
reproduction of the form of the induced
Chern Simons action on the lattice
shown to work in 3D by Coste and
L{\" u}scher [19] earlier). Then Kaplan
made the striking observation that a lattice chiral
theory at the defect just fell in our lap.
For the first time we had a concrete lattice approach that
transparently distinguished anomalous from non--anomalous
situations.

Narayanan and I started working on the
subject attempting to see what could be
in common between the continuum Pauli--Villars
like approach of
Frolov--Salvnov and the lattice one of Kaplan.
We felt that if there was a way to regularize
chiral gauge theories it could not be dependent
on some very special trick, involving a d+1
dimensional field theory and a mass defect,
nor could it depend on special chracteristics of the
gauge group.
We concluded that the
essential ingredient was the infinite number
of flavors and proceeded
to implement one example of this by constructing what is
nowadays referred to as the ``overlap'' [20].

At the same time, various other lattice [21] and non--lattice [22]
workers were studying the specifics of Kaplan's
proposal. After some discussion it became clear that
one could not afford gauge fields that had an extra
component and extra dependence on the additional
coordinate. In this respect the other workers
came to the same conclusion as we did, namely that
the single place the extra dimension should enter was in
the arguments the fermions depended on, i.e the
extra dimension could be viewed as a flavor space.
However, what was not accepted by some of the other workers
was that one needed an infinite flavor space and one could
not settle for a finite one.
Once you have a hammer,
everything looks like a nail, so it did not
take long until the infinite number
of fermions was sacrificed for a model
that had two wall
defects and looked like a Yukawa model,
but now with more fields
than before and a special mass matrix [23].
After some additional time it was declared that
the new model (known as the wave--guide model)
did not produce a chiral
theory after all [24]. The basic message that came
from this work was that Kaplan's proposal, once
implemented in the finite flavor version, would be
just another addition to the depressing list of failed
attempts. It would be a shame if this attitude became widely
adopted, and my main purpose in this talk is to
try to prevent this from happening.
Therefore, I'll get back to discussing [24] and related
work towards the end of my talk.

Let me start my more detailed discussion with
't Hooft's recent short paper: Let us denote
the left handed Weyl operator in a gauge background by $W_L (A)$.
The eigenvalues of $W_L (A)$ are gauge invariant, but not Lorentz
invariant. To avoid breaking Lorentz invariance by a continuum
regularization that relies on the eigenvalues
one sacrifices gauge invariance and considers the Dirac operator
$D(A)=W_R (0) + W_L (A)$ instead [25]. If $A$ has zero topological charge
$D(A)$ has an eigenvalue problem with Lorentz invariant but gauge
dependent eigenvalues. Exploiting the eigenvalues of $D(A)$ one
can regulate its determinant by adding a few Pauli Villars
Dirac fields with action $\bar\phi_i  [D(A)+ M_i ]\phi_i$. Since
all operators have the same eigenfunctions, the total
determinant
(resulting from the integration over all the original
fermion fields and the Pauli Villars regulator fields associated with
them) can be defined as the product over all the eigenvalues
combined mode by mode as dictated by the statistics of the respective
field. The infinite product over
the common modes can then be controlled
by choosing the PV fields and the masses $M_i$ in the usual manner.
't Hooft proves that the infinite product, with finite $M_i$'s, has a
finite limit for all $A$'s satisfying a mild requirement of pointwise
boundedness. Let us call this result $\det_{\rm reg} (A)$.
Consider now the expansion of $\log(\det_{\rm reg} (A))$ up to fourth
power in $A$ (in four dimensions). These terms collected in
$\Gamma_{\rm div} (A)$ will contain the very
familiar UV divergent diagrams. Assuming that anomalies cancel we know that
there exist local functionals of $A$ (potentially up to fourth power in $A$)
and dependent on the regulators $M_i$ called counter-terms,
$\Gamma_{\rm ct} (A)$,  such that $\Gamma_{\rm div} (A)+\Gamma_{\rm ct} (A)$
has a finite limit as $M_i\to\infty$. Consider therefore now
$\det_{\rm reg} (A)* \exp (\Gamma_{\rm ct} (A)) $$\equiv$
$\det^{\rm tr}_{\rm reg} (A)* \exp (\Gamma_{\rm ct} (A)+\Gamma_{\rm div} (A))$.
The $M_i\to\infty$ limit
of the term in the last exponent exists and is well known.
All that one needs to do now is to prove that the $M_i\to\infty$ limit
of the truncated determinant exists. Write
$\det^{\rm tr}_{\rm reg} (A)\equiv\det^{\rm tr}_{\rm bare} (A)*
\exp (\Gamma^{\rm tr}_{M_i} (A))$ where
$\Gamma^{\rm tr}_{M_i} (A)$ denotes the sum over all diagrams
with five or more external legs and with
only massive PV fields in the loop. It is a trivial matter to show that
this sum converges absolutely by bounding the Feynman integrands order
by order, exploiting their massiveness.
Moreover, the infinite $M_i$ limit
of the sum trivially vanishes. There is no $M_i$ dependence left
in $\det^{\rm tr}_{\rm bare} (A)$ (which represents the
sum of all original UV convergent diagrams). So
$\det^{\rm tr}_{\rm bare} (A)$
has an $M_i\to\infty$ limit and this concludes the
construction of the chiral determinant at zero topology.

One may ask why does one not simply deal with the perturbative sum over
the UV convergent original Feynman diagrams directly and why one needed
the eigenvalue analysis.
(Clearly, the main point is that as long
as we deal only with the determinant we have only
one diagram at each order $n$, nothing
like the prohibitive $n!$ one
always sees when investigating a full field theory.)
The reason has to do with the masslessness
of the original fields. The bound on the Feynman diagrams will
necessarily be measured in units of the physical size
of the system and therefore the sum would converge only for $A$'s of the
order of the inverse of this size. This set of $A$'s
has no chance of including all continuum backgrounds the lattice
would induce.
The limitation on the convergence of the
sum of perturbation theory can be easily understood directly:
we are exposed to divergences resulting from taking the
logarithm of a vanishing eigenvalue of $D(A)$. (This does not imply that
there is any problem with the product which is all we should care about.)
With an $A$
of the order of the inverse system size
one can easily arrange for an
exactly vanishing eigenvalue, for any boundary conditions
one works with. We conclude that our
estimate of the range of $A$'s for which all
UV--convergent Feynman diagrams can be summed was realistic. There is no
similar problem for the ultra--massive regulators and hence the
``summing of perturbation theory'' route works for a reasonable
range of $A$'s. In summary, 't Hooft uses non--perturbative means
to construct the sum over all UV convergent one (fermion--) loop
diagrams and this is the single piece of the chiral determinant that
cannot be handled by perturbation theory alone.

It is quite obvious that one needs a reasonable large range of $A$'s
for which the Grassmann integral is nonperturbatively defined.
't Hooft shows that any collection of lattice links can be mapped
into a continuum connection that obeys the bound needed for his
construction of the chiral determinant to work. Since the latter
is gauge invariant under continuum gauge transformations of $A$
(no anomalies) only the gauge invariant content of the $A$ gauge
configuration needs to be
related to the lattice configuration.
Only lattice gauge invariant information should affect the
gauge invariant content of the continuum connection and this
means one can apply convenient gauge transformations on the lattice,
during the construction of the mapping, as long as the
continuum gauge invariant data is unaffected. This is exploited
by 't Hooft who
gives a relatively simple construction, simplex by
simplex. There is an earlier construction (by G{\" o}ckeler et. al. [12])
of such a mapping that
avoids making any gauge choices, in the sense that
lattice gauge transformations get mapped consistently into continuum
gauge transformations. This is achieved by replacing standard patchwise
bundle definitions by possibly singular functions $A_\mu (x)$. The
singularities are physically unobservable. To preserve locality, the
number of singularities by far exceeds the minimal number
needed by gauge topology, but this is unavoidable.\footnote{*}{I am grateful
to Meinulf G{\" o}ckeler for
helping me understand the construction after the conference.}
Even for zero topology the ``functions'' $A_\mu (x)$ seem too singular
for 't Hooft's pointwise bound, but this can be remedied by singular
continuum gauge transformations.
This appears to defeat the main purpose of
maintaining simultaneously
explicit gauge invariance and locality throughout
the construction; finding the
appropriate global singular gauge transformation that gets rid of
all the singularities
(which is possible if the total topology is trivial)
might be a hard nonlocal problem. The nonlocality will not feed into
the total chiral determinant because the latter is gauge invariant;
therefore these difficulties are not a serious threat.
Note however that the individual eigenvalues 't Hooft employs are gauge
dependent, therefore at the technical level the
above comments are relevant.
Let me stress that I don't see anything fundamentally wrong
with picking convenient gauges from the beginning, as 't Hooft does.

A more serious issue concerning 't Hooft's construction is
encountered at non--zero topology. Assume
that the lattice gauge
field maps to a continuum gauge field given by a connection
on a nontrivial bundle. We do get the correct result,
namely, that the chiral determinant is zero,
but this happens essentially by definition.
Suppose however that we wish to compute the
expectation value of a 't Hooft vertex,
which gets contributions only from
nontrivial bundles. We no longer can use the eigenvalues of $D(A)$ to
proceed since the $W_R (0)$ part of it maps righthanded sections
of a trivial bundle
into left handed sections of a trivial bundle and the $W_L (A)$ part of
it maps left handed sections of a nontrivial bundle into right handed
sections of the same nontrivial bundle. Thus $D(A)$ maps objects of
one type into objects of a distinct type and therefore there
is no meaning to the concept of an eigenfunction with nonzero
eigenvalue. This is so on a torus, or any other compact base manifold.
(If one works on $R^4$ the situation is less clear to me. I doubt
however that it is wise to go to the infinite volume
limit immediately, without preceeding this step by
taking the UV limit.)
In summary, 't Hooft's approach has the potential of leading us to
an answer on the question of principle regarding the existence
of the chiral field theories we are discussing, but more work is
needed, even beyond technicalities. I am more doubtful
about this approach becoming useful
for carrying out practical non--perturbative computations
in the near future.
Let me stress again however that, in my view,
't Hooft does address the main problems faced by attempts to
define chiral theories and solves them by keeping the number of
fermions infinite, even when the theory is regulated.

Even when all is said and done in 't Hooft's approach one
still will need a discretization in order to compute
enough eigenvalues of $D(A)$ sufficiently accurately
to be able to estimate the infinite product reliably. It is clear
that a much preferable situation would be
to get the chiral determinant defined in terms of
a discretization that is identical to the one of the original lattice
gauge links. The overlap achieves this. Moreover, the overlap
is a more complete construction, in the sense that
nontrivial topology naturally fits in, and one has already a working
definition of regularized 't Hooft operators.
If we are willing to accept that there is no
danger in fixing the gauge, I have no doubt that the overlap
works correctly. Most of the efforts going into this
formalism at the present are to show that one
can avoid gauge fixing by
gauge averaging which would be by far more elegant.
It is unclear to me how the
``gauge covariant'' map of lattice gauge fields to
continuum singular ones [12]
could be implemented in practice, but if one had a way,
one could use it only to refine
the lattice on which the fermions live and
there is no real doubt that the overlap would work then.
I hope that the refining is avoidable employing gauge averaging
but this has not been proven yet. Let me stress however that this issue
has little to do with chiral fermions; that problem is
solved once one uses (in a controlled way) an infinite
number of fermions in the regulated system.

Let me turn now to the overlap and sketch its main properties.
The most physical starting point is to consider a vector theory with
several flavors of fermions.
$
L_{\psi} =i\bar\psi\sla D \psi +\bar\psi (M {{1+\gamma_5}\over 2} +
M^\dagger {{1-\gamma_5}\over 2} ) \psi$.
The Dirac equation can be written as
two coupled Weyl equations and we see that
there are $n_R = dim(ker M)$ massless righthanded fermions
and $n_L = dim (ker (M^\dagger ))$
left handed fermions.
As long as the number of flavors is finite, $M$ and $M^\dagger$
have equal rank and $n_R = n_L$. We cannot obtain a chiral theory,
regardless of how much we tune the mass matrices.

We therefore make flavor space into an infinite dimensional Hilbert
space and endow the operator $M$ on that space with an analytical
index equal to unity. We take flavor space as the space of all square
integrable functions on an (infinite) open ended line
parametrized by $s$. In the physical quantum theory the only single fermion
particle states are those whose wavefunctions are
normalizable also in the
$s$ direction. Actually,
real space is compactified, and, since we work
exclusively in Euclidean space, also time will be
compactified (and imaginary).
Naively, the theory so defined is unitary.

Assume that $M^\dagger M$ has a single zero eigenvalue and $MM^\dagger$ has
its lowest eigenvalue at a finite separation from zero. Such a
structure is stable against small deformations of $M$ (deformations
that change the spectrum of $M^\dagger M$ by only a limited amount).
Therefore, the device
we have found for making the theory chiral is stable
against radiative corrections. The observation that an internal
supersymmetric quantum mechanics was operative in flavor space
was our starting point in [9]. It was well known at that time
that so called ``second order formulations'' (forgotten by
now) failed because they did not have a mechanism protecting
the tree level vanishing of mass terms. The operator $M$ had
an index and this preserved the supersymmetry against
not too violent perturbations. One cannot realize an operator
with an index on a finite dimensional Hilbert space, since
the rank of a square matrix is always equal to the rank of its
hermitian adjoint. The correct way to think about an operator
with an index is to view it as a rectangular matrix. But, if we
did that, we accomplished nothing, since the vectorial formal
appearance of $L_\psi$ is lost and rather than one flavor space
we have two, one for the righthanded
fermions and another for the lefthanded ones.
If we give up on having an index we might try to decouple the
extra massless Weyl fermion from the gauge fields, but since
there is no protection against radiative corrections, the
wanted and unwanted massless Weyl fermions at tree level
mix by mass terms due to quantum fluctuations. This was tried
before the new proposals and with no success.
While one may hope to tune these undesirable perturbative
effects away the trouble one encounters with gauge field
topology is a strong hint that one is on the wrong track.

Keeping flavor space infinite even after UV
regularization makes the
Grassmann path integral sufficiently indefinite to violate
Ward identities
derivable by changes of variables involving all fermion flavors. Therefore,
gauge breaking effects are permitted even if the UV regularization is gauge
invariant. To separate the physical massless degrees of freedom from
the rest we scale $M$ by an auxiliary UV cutoff, $\Lambda$.

We now pick a reasonable realization of
$M$ with the objective of being
able to interpret the Grassmann integral in a useful way that will
lead us to well defined, completely regulated expressions for all
fermionic Green's functions in a fixed gauge background. The choice we
advocate at present is derived from Kaplan's proposal, known as the
wall choice: ${M\over\Lambda}=\partial_s + F^\prime (s) \equiv
e^{-F(s)}\partial_s e^{F(s)}$ with $F^\prime(s)=\epsilon (s)$.\footnote{*}{This
choice has the advantage of connecting us to the
Callan and Harvey physical picture for how charge
nonconservation operates. But, other choices for $M$ and
the space it acts on are also possible (even if we stick
to Callan and Harvey's examples,
we have the option to make flavor space the space of square
integrable functions on a two dimensional manifold [9] instead
of a one dimensional one).
The success of the basic approach of having an $M$ with an
index is not contingent on the outcome of the investigation of
the wall version. The paper by Frolov and Salvnov is some
limited evidence for this.}

The integral over the $\bar\psi_s , \psi_s$ for $s>0$ is interpreted as
describing a Hamiltonian evolution in the Euclidean time $s$ from
infinity to zero with a Hamiltonian $H^+$. Since there is homogeneity
in $s$ for $s>0$ $H^+$ is $s$-independent. Similarly, for $s<0$ we
have an $H^-$. Thus the chiral determinant is replaced by the overlap
${}_A\!\!<-|+>_A$ where the states $|\pm >_A$ are the ground states of
$H^\pm (A)$. The hermitian operators $H^\pm (A)$ depend parametrically
on the gauge field $A$ and have the structure of a system of noninteracting
fermions: $H^\pm (A) = a^\dagger h^\pm (A) a$ where $a$ is a fermion
annihilation operator (Dirac, group and space-time indices have been
suppressed). It is easy to read off $h^\pm$ from the path integral:
$
h^\pm = \pmatrix{ \pm m& W_L (A)\cr W_R (A) & \mp m }$.
$h^\pm$ are hermitian by $W_R (A) = W^\dagger_L (A)$.

The overlap does not define a function over the
space of $A$'s but rather
a U(1) bundle over that space (the
U(1) bundle can be taken down to gauge orbit space because the
$h^\pm (A)$ are gauge covariant). If we have a vector theory we
get the overlap factor from the lefthanded piece of the Dirac multiplet
and its complex conjugate from the right handed piece. This follows
by interchanging $M$ and $M^\dagger$ above. So, for the vector case we
do have a function and it is gauge invariant. In general, any gauge
breaking is restricted just to the phase of the overlap.

The overlap can vanish in two distinct ways:

1. ``Accidentally'' by having one
of the single particle wavefunctions in $|->$
orthogonal to the space spanned by the single
particle wavefunctions in $|+>$.
Accidental zeros in the overlap reflect
accidental zeros in the chiral
determinant. The simplest example is for free
chiral fermions on a torus with
periodic boundary conditions. A more interesting example
is the zero that occurs in the
interior of the discs Alvarez--Gaume and
Ginsparg embed into $A$ space [25].
That zero, although of topological origin
occurs for gauge fields of zero
topological charge and induces a winding of the
phase of the chiral determinant around the boundary of the disc.
Since the boundary of the disc is entirely on a single gauge orbit
this implies anomalies and explains in yet another way their
irremovability. That the zero induces a winding simply means that
in this respect the chiral determinant behaves as a generic complex
valued function of two real variables spanning the disc. It is quite
likely that the overlap will also behave generically in this sense.
This would not have been necessarily the
case had we insisted on separate
regularizations of the real and of the
imaginary part of the effective action.

2. The overlap can also vanish in a way that is
robust under small deformations
of the gauge background: Both hamiltonians preserve the fermion number
operator $\sum a^\dagger a$ so the ground states have definite such numbers.
When the numbers of negative
energy states in the two Dirac seas ($\pm$)
are different one gets a robust
zero and the strange but welcome effect
that the ``expectation value''
of an $a$ for example can be nonzero.
It is easy to see that this will
happen when the background carries
nonzero topology and use is made
of the known zero modes. Thus the overlap
has the ability to define
't Hooft vertices with the expected properties.

This leads us to the question how are fermionic
Green's functions defined.
In the domain wall picture the chiral states are wavepackets
localized in $s$ around $s=0$ which travel with the velocity
of light in the space directions.
Therefore $\bar\psi_s$ and $\psi_s$ at $s=0$
are reasonable interpolating fields for the physical fermions.
Hence the correspondence between
$<\bar\psi_L ...\bar\psi_L \psi_L ...\psi_L >_A$
and ${}_A\!\!<-|a^\dagger_L ....a^\dagger_L a_L .... a_L |+>_A$
which is compatible with our identification of the 't Hooft vertex.

As a result, for $A$ of trivial
topology (and a similar statement holds for
other topologies) we have that the
overlap representative of
$
{{<\bar\psi_L ...\bar\psi_L \psi_L ...\psi_L >_{A^g}}\over {<1>_{A^g}}}$
has a naive dependence on the gauge
transformation $g(x)$. This is
a desirable result since Fujikawa [26] has shown that anomalies
can be interpreted as a property of the
Grassmann integration measure, hence
anomalous behavior cancels out between the
numerator and the denominator of the expression above.
In the overlap this is a result of the fact that
a gauge transformation $g$ is represented in the $a$ Fock space by
$G(g)$ with $G(g)^\dagger a G(g) = g a$ and
$|\pm >_{A^g} = e^{i\phi(A,g)} G(g) |\pm >_A$.

A simple choice for the phase
(a section of the U(1) bundle over $A$ space) was
shown to reproduce various anomalies\footnote{*}
{While for perturbative backgrounds there exist
several derivations, when the background carries
nontrivial topology and one computes the anomaly in
the presence of an inserted 't Hooft vertex, only
the numerical check in the last paper in [20] is
available at present.} by
us and by Randjbar-Daemi and Strathdee.
(The latter authors also worked
out analytically some additional properties of the overlap;
I refer you to their papers for details [27].)
This choice for the phase
is referred to as the Wigner Brillouin choice because Wigner
Brillouin Schr{\" o}dinger perturbation theory is formulated
with the same phase choice.
The well defined states with this phase are
denoted by $|\pm >^{WB}_A$ and
satisfy ${}_0\!\!<\pm |\pm >^{WB}_A >0$
for some specified choice of the free
state. (For nontrivial topology
an equivalent definition can be chosen; all sectors can be
combined in a more general definition
which boils down to the above.)

Until now we have not yet introduced the lattice.
This shows that the overlap is something quite robust and meaningful
independently of the regularization.
By design, the overlap is supposed to make
regularization easy. Indeed, when putting the theory on the
lattice, what was once viewed as
the formidable doubling phenomenon
and considered to identify the heart
of the problem now appears
as a minor technical complication that
can be dealt with just as easily
as one deals with it when latticizing a
vector theory with massive Dirac
fermions:

The spinorial structure of $h^\pm$ is:
$
h^\pm =\pmatrix {B^\pm & C \cr C^\dagger & -B^{\pm}}$.
Before regularization one has $[B^\pm , C] =0$. In that case,
if one ignores infinities one easily
proves that ${}_A\!\!<-|+>_A \propto
\det C$ for a good global phase choice. But,
once one is on the lattice,
$h$ will become a finite matrix, and if also
$C$ is one, it would
be of square shape and would not be able to represent the continuum
infinite dimensional Weyl operator it
replaces because it will never
carry an index. Anyhow, if we could replace
faithfully the Weyl operator
by a square matrix, there would have
been no chiral fermion problem to start with.
(It is easy to check that
if one stubbornly nevertheless tries the
system protects itself against
our foolishness by producing doublers.)
So we need to break the
tight relationship between the overlap
and the determinant of $C$.

To make the overlap really different from
$\det C$ but still close
to it when restricted to a subspace
of slowly varying spinorial fields, we need to make the commutator
$[B^\pm , C] $ nonzero.
This invalidates the exact relation between
the overlap and the
determinant of $C$. Since we still want
$B^\pm \sim \pm m$ for
small momenta the simplest choice for $B$ is
the Wilson covariant mass term.
In the continuum, this choice
preserved both Lorentz and gauge covariance. This is all one does
when going to the lattice; the replacement of $C$ is the most naive
one. Of course, cutoff effects can be reduced by better choices.

One can prove that for any set of link variables $h^-$
always has
an equal number of positive and negative eigenvalues. This
is not true of $h^+$ who can have imbalances. By numerical checks
it is established that when the background
has nontrivial topology the right deficits occur.

What about gauge invariance ? On a lattice gauge invariance is always
attainable by gauge averaging. The relevant question is whether
gauge averaging introduces new nonlocal terms. Of course, if there
is gauge invariance even before gauge averaging, gauge
averaging induces no terms at all. But, clearly
(Forster, Nielsen, Ninomiya [28]),
one does not need exact gauge invariance for gauge averaging to
induce only local unimportant terms. If the theory is
anomalous, gauge averaging very likely does induce extra nonlocal terms
since what one is computing is
the partition function of a gauged Wess -- Zumino action in a fixed
gauge background. (Therefore,
once we make everything concrete we also have an explicit
proposal for trying to quantize an anomalous theory as first
suggested by Faddeev and Shatashvili [29]. Of course, it
may easily happen that we end up without a continuum limit.)
The WB phase choice induces a lattice version of
the WZ action that preserves several of the properties of the
continuum with respect to some discrete symmetries. The result is that,
for link variables $U$ and gauge transformation elements $g(x)$ close
to unity, the lattice action, when expanded in derivatives
has the WZ action as its leading term. However, there are also latticy,
subleading terms. When anomalies cancel the WZ term
disappears and one is left
with only the subleading terms. At the moment it is a conjecture that
gauge averaging in this case will be harmless, just adding some local
irrelevant
gauge  invariant terms to the effective gauge field action.\footnote{*}
{The employment of gauge averaging to restore
gauge invariance in lattice chiral models is by no
means a new idea; see for example ref. [30].} Of course,
this might require more care in the choice of the lattice version, but,
let me stress, since we could work in a fixed gauge throughout and this
should be OK in principle albeit inelegant, the validity of the
conjecture is an issue separate from
the solubility of the chiral fermion
problem.\footnote{**}{It is sometimes
said that the gauge degrees of freedom
cannot be unimportant because they fluctuate strongly. Actually, the
gauge degrees of freedom fluctuate maximally when there is exact
gauge invariance. The conjecture is that in the anomaly free case
the gauge degrees of freedom keep fluctuating strongly, in the sense
of being correlated only over distances of a few lattice spacings.}

The first full implementation of the overlap in a toy model has been carried
out successfully [31]. Together with Vranas, Narayanan and I showed that
in an overlap--latticized massless Schwinger model with one or more
flavors the right expectation
value for 't Hooft vertices is obtained in the continuum limit. This
strongly indicates that for massless QCD the
overlap will likely reproduce
't Hooft's solution to the axial U(1) problem in a concrete, nonperturbative
calculational scheme.

My view is that once we know how to preserve axial global symmetries
{\it exactly} in a vector theory, chiral gauge theories will present
no additional fundamental difficulties. Thus, I am optimistic on
the overlap. All Yukawa attempts to date are failures already in
the strictly massless vectorial context; these attempts have taught
us embarrassingly little.

Let me discuss now the
Yukawa attempts and the criticisms of the overlap coming from
there.
The basic difference we have with the Yukawa camp is
the following:  We believe that the {\it generic} lattice Yukawa model
for regularizing chiral fermions {\it has} to fail
since it cannot reproduce instanton effects even though it
is defined for all gauge fields (so it always gives obviously
wrong results for a subset of gauge fields that
are probabilistically not insignificant). Therefore we think that
one cannot learn much from the detailed way particular models
in this class fail. The correct conclusion from these failures
should be that one has to deal correctly with topological effects.
The Yukawa people think that topology is immaterial since
the failure occurs even for topologically trivial (actually,
even at zero gauge field strength) backgrounds; according to them the
real culprit is the presence of
{\it any} amount of breaking of gauge invariance.
We think that some amounts of gauge breaking on the lattice are
irrelevant in the continuum; the generic Yukawa approach
fails because once a theory cannot be
what you want it to in
a certain subsector of field space, it has no choice but
becoming something undesired everywhere (a self--consistent
interacting field theory is a severely constrained system).
This dispute will get resolved with time.

There is one case that has been investigated in detail and
where we can compare: There our construction works perfectly for
precisely the backgrounds accused of destroying the
wave--guide Yukawa lattice models.
The case where all the {\it specific} claims
are made is 2D U(1) gauge theory
with numbers of identical chiral multiplets divisible by four:
In the background of the trivial gauge orbit
($U_\mu (x) = g^\dagger (x+\mu ) g (x)$)
the appropriate
contribution of overlaps was proven to be gauge invariant
($g(x) $ independent) in the
last section of our last NPB paper [20]. On the other hand,
considering the same orbit,
Yukawa people claim that the coupling of
the fermions to the gauge degrees of freedom has dramatic
consequences [32] on the matter
content. While the overlap prescription for fermionic
Green's functions correctly
reproduces the naively gauge transformed free
field answer the Yukawa camp claims that there are hidden
unwanted excitations that alter the picture.
Somehow these excitations completely decouple from
the real part of the effective action (since by now we all agree
that employing the overlap in a vector theory works
correctly) and
also from the imaginary part of the effective
action (since they couple through $g(x)$ and the complete
lattice effective action is $g(x)$ independent).
Nevertheless, the Yukawa camp insists that the extra excitations
imply the failure of the overlap, meaning that they do not decouple
from the ``good'' excitations that control the continuum
limit of the full dynamical theory.
For this particular case it seems
impossible that we are both right:
Either our analytical proof is wrong (we double--checked our proof
numerically) or the claim made by the Yukawa people is wrong.
This dispute should have been resolved by now.

Some confusing terminology has been introduced along the way:
Originally there was a ``wave--guide'' model [23]
and an ``overlap'' [20].
The ``wave--guide'' gave an effective action that broke
gauge invariance both in the imaginary part and in the real part.
The overlap gave an effective action that broke gauge invariance
only in the imaginary part, and,
in addition, had enough exact discrete
symmetries to make the leading operator in a derivative expansion
be given by the appropriate Wess--Zumino action.
The overlap was arrived at from a path integral; to get
the overlap from there one needed to subtract some infinite, but
completely gauge invariant and real terms (formally).
In our first paper [9] it was pointed out that one can view
this subtraction as the addition of some heavy, gauge
invariant, Pauli--Villars fields; these fields are distinct
in their role from the Pauli--Villars fields used by Frolov and
Slavnov [8] in that they do not regularize any UV
divergences. For the overlap,
the subtraction was needed since the terms were infinite.
There were no analogous
subtractions in the original ``wave--guide''
case since there the total number of fields was finite,
so one needed not bother with terms that reflected some ``heavies''
in particular since they did not break gauge invariance.
In the so called ``modified wave--guide'' [33] the subtractions
and the associated above mentioned Pauli--Villars fields were added.
For small perturbative fields, and in the limit
of an infinite number of spin 1/2 fields the ``modified wave--guide''
effective action is the same as that of the overlap.
The essentials of this relationship
were given in section 6 of
our long NPB paper [20] where we also pointed out
its limited value. The subsequently introduced ``modified
wave--guide'' [33]
points out the relationship, but omits the limitation.
This we have tried to correct in a short recent publication [34],
which merely repeated the relevant parts of our larger paper.

As I mentioned already, there is a significant
difference between the gauge breaking in the overlap
limit of the ``modified wave--guide''
and the gauge breaking in the original ``wave--guide''.
To make this very clear let me mention that if
one uses the ``wave--guide'' model to realize
each of the independent
Weyl components of the massless Dirac fermions in
a vector theory one will end up with a different theory
than intended, typically one with a doubled set of fermions.
On the other hand, if one uses the ``modifed wave--guide''
in the ``wave--guide's'' stead,
{\it and one restricts oneself artificially to the zero
topological sector}, due to the relation to the overlap,
no extra fermions would appear.
It is therefore important to note that
no serious analyses, numerical or otherwise,
of the so called ``modified wave--guide'' have been reported,
except the ones carried out for the overlap in our papers.
Our tests were successful as far as they went.
We shall have to wait until a complete simulation (including
dynamical gauge fields) is carried out, but the indications
are that it will also succeed.

In summary, to date, the critical comments
Yukawa people made about the overlap are either unsubstantiated
or provably wrong.
The carriers
of the Yukawa banner and the
message that no real progress has been made
in the field
are Golterman and Shamir. I refer you to their papers for more
details and
references. A reasonably up to date, earlier
review, reflecting a less inflammatory version of the
Yukawa point of view, has been authored by
Karl Jansen who is here.

I should comment on the necessity of having an infinite number of
fermions as opposed to its desirability. If the regulated
theory has a unique vacuum in finite volumes while
the fermions appear only bilinearly and if the maximal number
of nonanomalous U(1)'s is exactly preserved, I see no other way.
However, if one gives up on either of the above, there may be
other schemes. For example, my arguments do not prove that
the ROME II approach [2] (the one including Majorana mass terms)
has to fail. However, this particular approach
may have another problem, related to non--perturbative BRS invariance
(the problem appears in the vector context and has nothing to do with
chiral fermions, there is an old PLB article of mine on this [36]).
A discussion
would take us too far afield, and anyhow, I can't make a more
definite statement.

My final message is that the overlap proves that one can deal with the
infinite number of fermions and that the result is a rather elegant
construction. Rather than dwelling on whether it is absolutely
necessary or not it makes more sense to proceed to carry out
increasingly demanding checks and, eventually, try to get a first
stab at the question of nonperturbative existence of an
asymptotically free chiral gauge theory. It is important that, at least
until we get convinced on the question of principle, we have
a tool that is sufficiently close to continuum constructions that
we can plausibly attach a possible failure to obtain a good
continuum limit to the target theory rather than the regularization
method. So we should try to avoid as much as possible ``dirty''
methods, even if eventually, they would be the most practical. The
overlap is sufficiently practical that we don't have to worry about this
at the present stage. The first set of checks has to be (and is carried
out at present) in two dimensions. It is just too easily done
relatively to four dimensions in
numerical terms on the one hand and, on the other,
it is highly unlikely that a scheme
which fails in two dimensions would work in four. This is not to say
that in four dimensions new difficulties cannot appear; we just should
remember that all older attempts failed already in two dimensions.

\vskip 2cm
\leftline{\bf Acknowledgments}
\vskip  1cm
I would like to thank the organizers for inviting me to participate in
this stimulating workshop which reassured me that field theory isn't
quite dead yet.

All my work on the chiral problem was done in close collaboration
with R. Narayanan. In our more recent numerical work also P. Vranas
was heavily involved.
Both the verbal exposition and this write--up have benefited from
the privately shared insights of many people
(in random order): R. Narayanan, P. Vranas,
Y. Aharonov, M. Srednicki, G. 't Hooft,
M. L{\"u}scher, E. Witten, W. Bardeen, R. Levien,
S. Randjbar--Daemi, M. Golterman, Y. Shamir,
L. Alvarez--Gaume, D. Kaplan, T. Banks, M. G{\"o}ckeler,
R. Shankar, S. Shatashvili, A. Casher, I. Montvay, K. Jansen,
N. Seiberg, J. Smit, M. Testa, A. Slavnov, G. Schierholtz,
U.--J. Wiese, R. Brower, D. Weingarten. I am sure there are many
more whose names I forgot to mention; for this I apologize.
Of course, none of these
people should be held responsible for any wrong statements I have
made. I am equally certain that my list of references is
incomplete and misleading at one point or another; for this I also
apologize.

My research is supported in part by the DOE under grant number
\# DE--FG05--90ER40559.

\vfill\eject

\leftline{\bf References}
\vskip 1cm
\item{[1]} Y. Shamir, Proceedings of Lat95, Nucl. Phys. B.
(Proc. Suppl.), to appear; G. M{\"u}nster, Proceedings of Lat94,
Nucl. Phys. B (Proc. Suppl.)
42(95) 162; D. Petcher, Proceedings of Lat92, Nucl. Phys. B (Proc. Suppl.)
30(93) 50; I. Montvay, Proceedings of Lat91, Nucl. Phys. B (Proc. Suppl.)
26(91) 57; J. Shigemitsu, Proceedings of Lat90, Nucl. Phys. B (Proc. Suppl.)
20(90) 515; M. F. L. Golterman, Proceedings of Lat90,
Nucl. Phys. B (Proc. Suppl.) 20(90) 528.
\item{[2]}  L. Maiani, G. C. Rossi, M. Testa, Phys. Lett. B261 (1991) 479;
Phys. Lett. B292 (1992) 397; A. Borrelli, L. Maiani, R. Sisto, G. C. Rossi,
M. Testa, Nucl. Phys. B333 (1990) 335.
\item{[3]}  Footnote on p. 1173 of H.B. Nielsen, M. Ninomiya, Nucl. Phys.
B193 (1981) 173.
\item{[4]}  Footnote 1 on p. 4016 of T. Banks, A. Dabholkar,
Phys. Rev. D46, (1992) 4016.
\item{[5]}   H. B. Nielsen, S. E. Rugh,
Proceedings of the Rome Workshop on
Nonperturbative Aspects of Chiral Gauge
Theories, Nucl. Phys. B (Proc. Suppl.)
29B,C (1992) 200.
\item{[6]} D. Kaplan, Phys. Lett. B288 (1992) 342.
\item{[7]} C. G. Callan, J. Harvey, Nucl. Phys. B250 (1985) 427.
\item{[8]} Frolov, A. Slavnov, Phys. Lett. B309 (1993) 344.
\item{[9]} R. Narayanan, H. Neuberger, Phys. Lett. B302 (1993) 62.
\item{[10]} G. 't Hooft, Phys. Lett. B349 (1995) 491.
\item{[11]} R. Flume, D. Wyler, Phys. Lett. B108 (1982) 317.
\item{[12]} M. G{\"o}ckeler, G. Schierholz,
Proceedings of the Rome Workshop
on Nonperturbative Aspects of Chiral Gauge Theories, Nucl. Phys. B
(Proc. Suppl.) 29B,C (1992) 114. M. G{\"o}ckeler, A. S. Kronfeld,
G. Schierholz,
U.--J. Wiese, Nucl. Phys. B404 (1993) 839.
\item{[13]} H. Neuberger, M. Sc. Thesis, 1975, (PITT-86-09).
\item{[14]} H. B. Nielsen, M. Ninomiya, in Trieste Conference on
Topological
Methods in Quantum Field Theories, World Scientific, (1991) and
references therein.
\item{[15]} M. Atiyah, I. M. Singer, Ann. Math. 93 (1971) 139.
\item{[16]} G. 't Hooft, Phys. Rev. Lett. 37 (1976) 8;
Phys. Rev. D14 (1976) 3432.
\item{[17]} R.  Stora, in Cargese lectures, 1983; B. Zumino, in Relativity,
groups and topology II, Les Houches, 1983, eds. B. S. Dewitt and R. Stora
(North--Holland, Amsterdam, 1984).
\item{[18]} J. Goldstone, F. Wilczek, Phys. Rev. Lett. 47 (1981) 986.
\item{[19]} A. Coste, L{\"u}scher, Nucl. Phys. B323 (1989) 631.
\item{[20]} R. Narayanan, H. Neuberger, Nucl. Phys. B412 (1994) 574,
Nucl.Phys. B443 (1995) 305.
\item{[21]} M. F. L. Golterman, K. Jansen, D. B. Kaplan,
Phys. Lett. B301 (1993) 219; K. Jansen, Phys. Lett. B288 (1992) 348;
Y. Shamir, Nucl. Phys. B406 (1993) 90, Nucl. Phys. B417 (1993) 167,
Phys. Lett. B305 (1992) 357. C. P. Korthas-Altes, S. Nicolis, J. Prades,
Phys. Lett. B316 (1993) 339.
\item{[22]} J. Distler, S.--Y. Rey, hep-lat \# 9305026, PUPT--1386.
\item{[23]} M. F. L. Golterman, K. Jansen, D. N. Petcher, J. Vink,
Phys. Rev. D 49 (1994) 1606.
\item{[24]} M. F. L. Golterman, Y. Shamir, Phys. Rev. D 51 (1995) 3026.
\item{[25]} L. Alvarez--Gaume, P. Ginsparg, Nucl. Phys. B 243 (1984) 449.
\item{[26]} K. Fujikawa, Phys. Rev. Lett. 42 (1979) 1195.
\item{[27]} S. Randjbar--Daemi, J. Strathdee, Phys. Rev. D 51 (1995) 6617,
Nucl. Phys. B 443 (1995) 386, Phys. Lett. B348 (1995) 543, hep-th \# 9510067
IC/95/305.
\item{[28]} D. Foerster, H. B. Nielsen, M. Ninomyia,
Phys. Lett. B94 (1980) 135.
\item{[29]} L. D. Faddeev, S. Shatashvili,
Phys. Lett. B167 (1986) 225,
Theor. Math. Phys. 60 (1984) 206.
\item{[30]} J. Smit, Proceedings of the Rome Workshop
on Nonperturbative Aspects of Chiral Gauge Theories,
Nucl. Phys. B (Proc. Suppl.) 29B,C (1992) 83.
\item{[31]} R. Narayanan, H. Neuberger, P. Vranas,
Phys. Lett. B 353 (1995) 507.
\item{[32]} M. F. L. Golterman, Y. Shamir,
Proceedings of Lat95, Nucl. Phys.
B. (Proc. Suppl.), to appear, hep-lat \# 9509027.
\item{[33]} M. F. L. Golterman, Y. Shamir, Phys. Lett. B353 (1995) 84.
\item{[34]} R. Narayanan, H. Neuberger, Phys. Lett. B358 (1995) 303.
\item{[35]} K. Jansen, DESY--94--188 hep-lat \# 9411016.
\item{[36]} H. Neuberger, Phys. Lett. B183 (1986) 337.

\vfill\eject\end